\documentclass[conference]{IEEEtran}
\IEEEoverridecommandlockouts
\usepackage{cite}
\usepackage{amsmath,amssymb,amsfonts}
\usepackage{graphicx}
\usepackage{textcomp, multirow}
\usepackage{xcolor, soul, cancel}
\usepackage{booktabs}
\usepackage{longtable}
\usepackage{algorithm2e}
\usepackage{hyperref}
\usepackage{balance}
\usepackage{tabularx}
\usepackage[noend]{algorithmic}

\SetKwComment{Comment}{/* }{ */}

\def\BibTeX{{\rm B\kern-.05em{\sc i\kern-.025em b}\kern-.08em
    T\kern-.1667em\lower.7ex\hbox{E}\kern-.125emX}}

\begin{document}

\title{Optimizing Fine-Grained Parallelism Through Dynamic Load Balancing on Multi-Socket Many-Core Systems
}

\makeatletter
\newcommand{\linebreakand}{%
  \end{@IEEEauthorhalign}
  \hfill\mbox{}\par
  \mbox{}\hfill\begin{@IEEEauthorhalign}
}
\makeatother

% \author{
% \IEEEauthorblockN{1\textsuperscript{st} Wenyi Wang}
% \IEEEauthorblockA{\textit{Department of Computer Science} \\
% \textit{The University of Chicago}\\
% Chicago, USA \\
% wenyiw@uchicago.edu}
% \and
% \IEEEauthorblockN{2\textsuperscript{nd} Maxime Gonthier}
% \IEEEauthorblockA{\textit{Department of Computer Science} \\
% \textit{The University of Chicago}\\
% Chicago, USA \\
% mgonthier@uchicago.edu }
% \and
% \IEEEauthorblockN{3\textsuperscript{rd} Poornima Nookala}
% \IEEEauthorblockA{\textit{Intel}\\
% Chicago, USA \\
% nookala.poornima@gmail.com}
% \and
% \linebreakand \IEEEauthorblockN{4\textsuperscript{th} Haochen Pan}
% \IEEEauthorblockA{\textit{Department of Computer Science}\\
% \textit{The University of Chicago}\\
% Chicago, USA \\
% haochenpan@uchicago.edu}
% \and
% \IEEEauthorblockN{5\textsuperscript{th} Ian Foster}
% \IEEEauthorblockA{\textit{Department of Computer Science} \\
% \textit{The University of Chicago}\\
% Chicago, USA \\
% foster@uchicago.edu}
% \linebreakand \IEEEauthorblockN{6\textsuperscript{th} Ioan Raicu}
% \IEEEauthorblockA{\textit{Department of Computer Science} \\
% \textit{Illinois Institute of Technology}\\
% Chicago, USA \\
% iraicu@cs.iit.edu}
% \and 
% \IEEEauthorblockN{7\textsuperscript{th} Kyle Chard}
% \IEEEauthorblockA{\textit{Department of Computer Science} \\
% \textit{The University of Chicago}\\
% Chicago, USA \\
% chard@uchicago.edu}
% }
\author{
Wenyi Wang\textsuperscript{1}, Maxime Gonthier\textsuperscript{1}, Poornima Nookala\textsuperscript{2}, Haochen Pan\textsuperscript{1}, Ian Foster\textsuperscript{1}, Ioan Raicu\textsuperscript{3}, Kyle Chard\textsuperscript{1}
\linebreakand \textsuperscript{1}\textit{Department of Computer Science, The University of Chicago, Chicago, Illinois, USA}
\linebreakand 
\textsuperscript{2}\textit{Intel, Chicago, Illinois, USA}
\linebreakand 
\textsuperscript{3}\textit{Department of Computer Science, Illinois Institute of Technology, Chicago, Illinois, USA}
\linebreakand wenyiw@uchicago.edu, mgonthier@uchicago.edu, nookala.poornima@gmail.com, 
\linebreakand haochenpan@uchicago.edu, foster@uchicago.edu, iraicu@iit.edu, chard@uchicago.edu
}

\maketitle
% ABSTRACT

\begin{abstract}
Achieving efficient task parallelism on many-core architectures is an important challenge.
The widely used GNU OpenMP implementation of the popular OpenMP parallel programming model incurs high overhead for fine-grained, short-running tasks due to time spent on runtime synchronization.
In this work, we introduce and analyze three key advances that collectively achieve significant performance gains.
First, we introduce XQueue, a lock-less concurrent queue implementation to replace GNU's priority task queue and remove the global task lock. 
Second, we develop a scalable, efficient, and hybrid lock-free/lock-less distributed tree barrier to address the high hardware synchronization overhead from GNU's centralized barrier.
Third, we develop two lock-less and NUMA-aware load balancing strategies.
We evaluate our implementation using Barcelona OpenMP Task Suite (BOTS) benchmarks.
We show that the use of XQueue and the distributed tree barrier can improve performance by up to 1522.8$\times$ compared to the original GNU OpenMP. We further show that lock-less load balancing can improve performance by up to 4$\times$ compared to GNU OpenMP using XQueue.
% Through a rich set of profiling and instrumentation tools, we are able to investigate the runtime behavior of GNU OpenMP and improve its performance on fine-grained tasks by many orders of magnitude. 
\end{abstract}


% INTRODUCTION
\section{Introduction}

The emergence of many-core computing systems with concurrency levels from hundreds on CPUs to thousands on GPUs has motivated the adoption of Non-Uniform Memory Access (NUMA) architectures, which offer asymmetric access to cache and memory banks.
Programming these systems using a shared memory programming model requires efficiently mapping threads to cores, which becomes increasingly challenging as the number of cores grows and task runtimes decrease. We focus here on addressing two significant challenges: 
first, overheads associated with use of hardware primitives to synchronize shared memory accesses across many cores and threads of execution;
and second, that shared memory programming models are not NUMA-aware. 

In the task parallel programming model,
computation is broken down into inter-dependent tasks that can be executed concurrently on various cores while respecting data dependencies.  
Various parallel languages and libraries support this model, such as OpenMP~\cite{miscname4}, Cilk~\cite{10.1145/209936.209958}, and Unified Parallel C (UPC)~\cite{upc}.
We focus on OpenMP, a task-centric model in which higher-level parallel constructs such as loops are translated into fine-granularity tasks with dependencies, which the runtime must dynamically schedule to available resources.
When a task is enabled by some thread, it is conceptually queued for execution by a future available thread.
Unfortunately, OpenMP implementations and their tasking data structures often scale poorly due to the excessive use of expensive synchronization operations, such as locks, to resolve dependencies~\cite{nookala_enabling_2021, navarro-torres_synchronization_2021, David_Guerraoui_Trigonakis_2013, morrison_scaling_2016}.

Two approaches to avoiding the performance degradation associated with locks are \textit{lock-free programming}, which typically relies on atomic hardware operations such as compare-and-swap to synchronize without locks, and \textit{lock-less programming}, which relies on methods to safely manipulate shared data without using locks.
We focus here on lock-less programming in which no atomic primitives are used.
In prior work we proposed the use of a lock-less, concurrent, multi-producer multi-consumer (MPMC) task queue, called XQueue, in the LLVM-based OpenMP (LOMP)~\cite{nookala_enabling_2021}. 
We showed significant performance improvements, often delivering 4--6$\times$ speedup, compared to native LLVM OpenMP by reducing the time spent on synchronization operations. 

In this paper, we focus on GNU OpenMP (GOMP)---the most common OpenMP implementation that is a part of the mainstream compiler infrastructure GNU Compiler Collection (GCC).
We specifically address the unscalable and entangled implementation of GNU OpenMP for fine-grain tasks, overcoming restrictive synchronization such as the excessive use of locks.
We then propose, implement, and evaluate two lock-less NUMA-aware, dynamic load balancing (DLB) algorithms.  
The main contributions of this work are as follows:

\begin{enumerate}
    \item We integrate XQueue, a lock-less, relaxed order MPMC task queue, with GOMP, replacing GNU's unscalable priority task queue and global task lock.
    Our benchmark results show that XQueue-enhanced GNU OpenMP (XGOMP) achieves up to 96.5$\times$ performance improvement (for an NQueens application) compared to the original GOMP.
    
    \item We propose a new hybrid lock-free (gathering)/lock-less (releasing) distributed tree barrier that yields a theoretical lower bound of half the atomic memory access operations than GNU's existing barrier which relies on a globally shared atomic task counter. Our approach differs from prior work in LOMP that uses LLVM’s default lock-free (not lock-less) barrier.
    We show that this optimization (XGOMPTB) improves the performance by up to 15.8$\times$ (NQueens) compared to XGOMP, and up to 1522.8$\times$ (NQueens) compared to GOMP.

    \item To the best of our knowledge, we propose the first lock-less, NUMA-aware dynamic load balancing algorithms for OpenMP tasking. Our approach significantly mitigates load imbalance for both fine-grained and coarse-grained parallelism and achieves significant performance improvements over all previous work. 
    %Our approach differs from our prior work that relies on a static load balancing (SLB) approach. 
    We also present a comprehensive study of configuration parameters that shows that all benchmarks considered achieve better performance compared to XGOMPTB. 
 
    \item We design and implement new software profiling tools for GNU to study and optimize how hardware, application, and OpenMP characteristics influence tasking performance.  
    %that improve on mainstream profiling tools such as Intel VTune and Scalasca.
    We use these tools to collect statistics such as time spent on task creation, task queue operations, task execution, and per-thread task locality, providing a deeper understanding of performance behavior.
    We learn that data and task locality plays a critical role in affecting application performance.

    \item We present guidelines for practitioners based on studies of our lock-less DLB strategies. We explore the relationship between task size, steal size, and performance, and analyze memory and cache behavior.

\end{enumerate}

The rest of this paper is organized as follows. 
\autoref{sec:motivation-and-background} introduces the motivation and background of this work. \autoref{sec:gnu-xtask} describes our integration of XQueue in GNU OpenMP and the design of our distributed tree barrier. % and the design of our GNU-XTask. 
\autoref{sec:dlb} describes our lock-less dynamic load balancing strategies.
\autoref{sec:software-profiling-tools} outlines how we instrument our code to measure performance. 
\autoref{sec:eval} presents a performance evaluation of our work. %GNU OpenMP variant and GNU-XTask.
\autoref{sec:application} presents a real-world application that benefits from our work.
\autoref{sec:performance-tunning} discusses how people should tune our DLB for optimal performance.
\autoref{sec:related-work} reviews related work.
\autoref{sec:discussion} summarizes our contributions.

%  MOTIVATION
\section{Motivation and Background}
\label{sec:motivation-and-background}

The open source GCC is the official compiler for GNU and the standard compiler for many Linux distributions. For OpenMP programs, GCC compiles and links programs with \textit{libgomp} (GOMP), which implements OpenMP standards but with GNU-specific customizations. LLVM also has an implementation of OpenMP and uses its own compiler, \textit{Clang}, %which is 
primarily designed for performance. 
The performance gap between the GNU and LLVM OpenMP implementations can be significant: see \autoref{fig:bots-gomp-vs-llvm}. 
Here we show execution times for various OpenMP applications from the Barcelona OpenMP Task Suite (BOTS) benchmark~\cite{duran_barcelona_2009} when using GNU OpenMP (GOMP), LLVM OpemMP (LOMP), and an implementation of XQueue in LLVM OpenMP (XLOMP). 
We run these experiments on an Intel Skylake-192 machine, with 192 cores (384 hardware threads). 
In some benchmarks, GOMP can be $>$1000$\times$ slower than LOMP and $>$4400$\times$ slower than XLOMP. %~\cite{nookala_enabling_2021}. 
We attribute the GOMP performance issue to its excessive use of synchronization primitives, including locks and atomic operations.

\begin{figure}[!ht]
    \centering
    \includegraphics[width=\columnwidth,trim=2.5mm 3mm 2mm 2mm,clip]{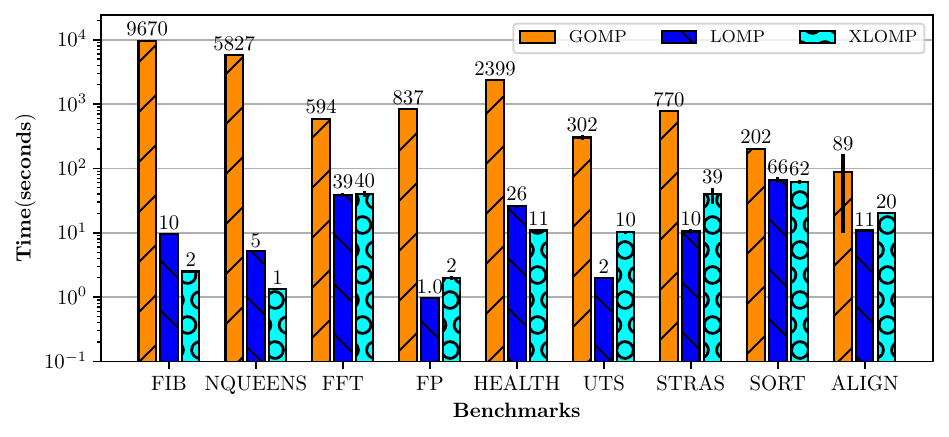}
    \caption{Execution times for nine BOTS benchmarks with three OpenMP implementations (GOMP, LOMP, and XLOMP) each using 192 threads.}
    \label{fig:bots-gomp-vs-llvm}
\end{figure}

\subsection{GNU OpenMP (GOMP)}
The GOMP runtime library handles OpenMP's parallel region as a \emph{team} of worker threads (i.e., workers). 
An OpenMP \emph{team} creates and manages a group of OpenMP threads. The team itself is often created under the OpenMP directive \emph{parallel}. 
OpenMP wraps platform-specific or user-defined threads. In our experiments, we use the  \emph{pthread}~\cite{pthread} library. 
GNU OpenMP manages tasks, for example, enqueuing and dequeuing tasks, using a single globally shared priority task queue and a child task queue for each task. 
The global priority task queue maintains the system-wide information, while the child queue maintains child tasks' dependencies and priorities. 
GNU OpenMP uses a single global task lock to protect critical regions for task management, scheduling, and other runtime bookkeeping operations. 
When a worker reaches a scheduling point, it will first acquire the lock before it enters the critical region. When there are no tasks to be executed or scheduled, and current tasks do not have any dependent tasks, the worker exits or waits for other workers to complete.

\subsection{XQueue}

XQueue~\cite{nookala_enabling_2021} is a lock-less MPMC, out-of-order queuing mechanism that can scale to hundreds of threads with little contention for hardware resources. XQueue uses B-queue, a concurrent Single-Producer-Single-Consumer (SPSC) lock-free queue designed for efficient core-to-core communication. The latency of B-queue operations can be as low as 20 cycles.

\begin{figure}[!ht]
    \centering
    \includegraphics[width=0.5\textwidth]{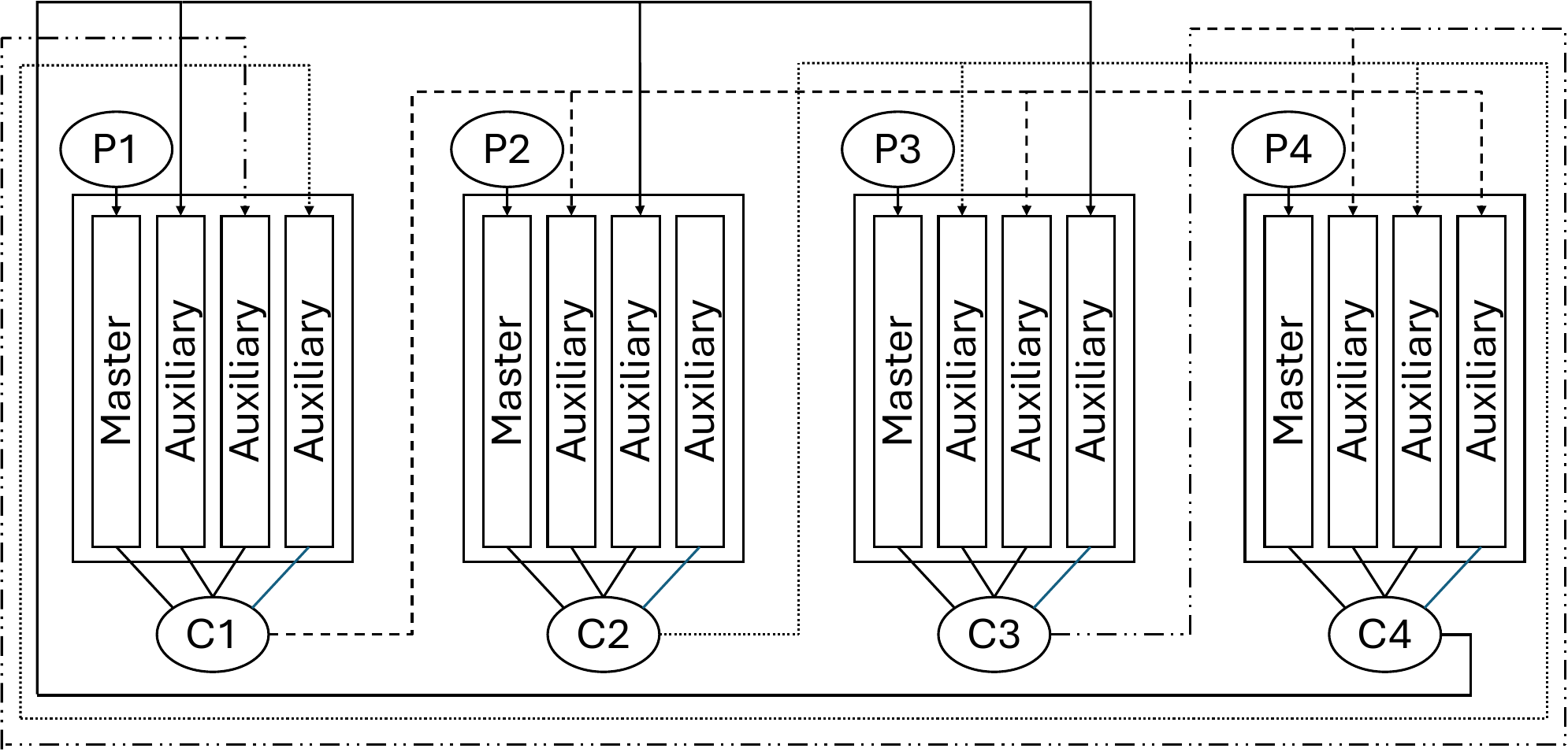}
    \caption{XQueue on a 4-core system. $C_{\text{i}}$ is consumer $i$, $P_{\text{i}}$ is producer $i$.}
    \label{fig:xqueue-design}
\end{figure}

\autoref{fig:xqueue-design} illustrates how XQueue is used on a 4-core system. 
%Each core has one worker thread with one SPSC queue for each other worker: one \emph{Master} queue and the others are \emph{Auxiliary} queues.
Each core has one worker thread with a single SPSC \emph{Master} queue and one \emph{Auxiliary} SPSC queue for each other worker.
Each item in a queue is a task pointer. Upon task creation, the worker enqueues the task to either its own master queue or to an auxiliary queue of another worker. It applies a round-robin approach across these queues starting with the master queue. If the chosen queue is full, the worker instead executes that task immediately. When dequeuing, the worker first dequeues from the master queue, if no task is found, it then dequeues a task from its auxiliary queues. The dequeued task is also executed immediately.

\subsection{Load Balancing and Work Stealing}
Integrating XQueue with OpenMP is likely to lead to load imbalance as the static load balancing (SLB) approach will assign tasks to threads without considering other tasks assigned to those threads. Thus, it is important to consider more advanced load balancing techniques to improve performance.

One popular load balancing technique is \textbf{work stealing}. 
A worker may be either a \textit{victim} or a \textit{thief}. 
A thief is an underloaded worker who tries to steal work from other overloaded workers. 
A worker being stolen from is a victim. 
Typically, \textit{pull-based work stealing} is used, in which a thief initiates and pulls a task from a victim. 
However, the queue from which tasks are stolen requires mutual exclusive access as it is accessed by both thief and victim.
As a result, traditional work stealing methods do not match our lock-less scenario due to their use of synchronization mechanisms, such as mutexes, spinlocks, and atomic operations, to ensure thread safety and correctness~\cite{NOOKALA2024444}.  % including atomic operations. 
Use of such methods would lead to the runtime wasting more cycles on cache invalidation, cache misses, and excessive main memory access. 

A second issue with existing work stealing techniques is that they are not NUMA-aware.
In a NUMA architecture, we can distinguish between data being either NUMA-local and NUMA-remote %data locality on a shared-memory NUMA architecture 
with respect to a core. 
Accessing NUMA-local data incurs less latency, and thus cores spend less time waiting for memory operations and can achieve higher throughput.
Work stealing algorithms must therefore consider whether a thief is to steal from a NUMA-local worker or a NUMA-remote worker, depending on application characteristics.

% GNU-XTASK
\section{Fine-Grained Parallelism in GOMP }
\label{sec:gnu-xtask}
Our goal is to transform GOMP into a high-performance runtime library that can scale to hundreds of threads and support extremely fine-grained tasks. To this end, we leverage the lock-less, concurrent \textit{XQueue} data structure and replace GOMP's centralized barrier with a distributed tree barrier.

\subsection{XGOMP: XQueue Integration with GOMP}
\label{sec:xgomp}

GOMP starts a \emph{parallel} region by first calling \texttt{gomp\_team\_start} and then \texttt{gomp\_thread\_start} to spawn a team of worker threads.
The master thread and the worker threads allocate XQueue's task queues inside \texttt{gomp\_team\_start} and \texttt{gomp\_thread\_start}, respectively, to ensure each thread has its task queue ready before it enters the thread dock
(the thread dock is a simple barrier that ensures that all worker threads have been properly initialized before the OpenMP tasking routine starts).

We modified \texttt{GOMP\_task} and \texttt{GOMP\_taskwait} to decouple GOMP tasking from the global task lock and priority queue. 
\texttt{GOMP\_task} is called when an explicit \emph{task} directive is encountered. 
It allocates a task and ensures its correct state before entering the critical region using the global task lock to enqueue this task to the priority queues.
With XQueue, we atomically update the parent task's dependency, push the task directly to the target queue,
and atomically increment the global task count, which is used by the centralized barrier.

Many of GOMP's runtime routines rely on variables that are protected by the global task lock. 
We removed the runtime dependencies of all those variables except the variable that tracks the global task count. 
The GOMP native tasking barrier uses this count as a termination signal. The count is updated when global task status changes, for example, it decrements when a task finishes. We convert this variable to an atomic variable with an acquire-release memory order strategy.

This phase of implementation removes GNU's significant lock contention for massively parallel, fine-grained tasks. However, it is possible that excessive use of atomic operations could still be a performance bottleneck. The atomic task count is deeply entangled with GNU's barrier routine, and therefore, we next replace it with a new barrier design. 

\subsection{XGOMPTB: Adding an Efficient Distributed Tree Barrier}
\label{sec:xgomptb}

There are several different barrier implementations in GOMP, all of which are implemented in a centralized manner. Here we specifically focus on the team barrier, which is usually implicitly placed at the end of a \emph{parallel} region during compilation and manages access to the global task count.
Updating the states of the team barrier requires acquisition of the global task lock. 
The centralized team barrier releases when the worker acquires the lock and meets the following conditions: 
i) it is the last worker entering the same barrier, and 
ii) the global task count is 0. 

In our XGOMP implementation, we modified the centralized team barrier such that it releases when each worker meets the following conditions: 
i) the global task count is 0; 
ii) current task of the worker has no dependencies;
and iii) all other workers have reached the same barrier.

We then replaced GNU's centralized team barrier with an efficient distributed tree barrier. We adopt a similar gather-release pattern as used in LLVM; however, unlike LLVM's lock-free barrier, we design a new hybrid barrier that performs lock-free gathering and lock-less releasing. Our approach yields a theoretical lower bound of half the atomic memory access operations.
Our barrier connects workers in a binary tree topology. 
A worker is gathered when: i) all workers have entered the barrier; ii) the worker is idle, finding no tasks to execute; iii) the worker's current task has no unfinished dependencies; and iv) all of its children workers' barriers are gathered. 
A gathered worker atomically updates the \textit{complete} flag of its parent. 
As this complete flag is only accessible by the child and parent, its hardware contention is low.
When the root worker is gathered, it then signals a release with a tree broadcast, lock-lessly updating the \textit{release} flag of each worker so that the worker can safely exit the barrier.

With these modifications, we have now removed all excessive atomic operations in GNU OpenMP. 
The tree barrier combines lock-free and lock-less techniques; it is also easy to understand and implement.
To the best of our knowledge, our approach is the first implementation of a hybrid distributed tree barrier for OpenMP tasking with carefully designed lock-less logic. A distributed tree barrier was briefly introduced to LLVM in 2021; however, it was later reverted due to unresolved errors. %At this time, there is no distributed barrier in LLVM. 
LLVM’s barrier implementation uses locks, atomic operations, and does not implement lock-less optimizations.

\section{Dynamic Load Balancing}
\label{sec:dlb}
XQueue uses a static round-robin load balancer to distribute tasks among workers. This approach has two limitations: first, load imbalance can occur, and second, data locality in NUMA machines is not considered, which may result in slower data access and poor performance. 

To address these limitations, we design a lock-less messaging protocol and propose two lock-less DLB strategies: NUMA-aware Redirect Push \textbf{(NA-RP)} and NUMA-aware Work Stealing \textbf{(NA-WS)}.
These strategies are the first lock-less DLB techniques for OpenMP tasking, with NUMA-awareness.
We first analyze load imbalance in our XGOMPTB implementation and then present our lock-less messaging protocol and the two DLB strategies.

\subsection{Measuring load imbalance}
\label{subsec:imbalance}
To understand the load imbalance, we collect and study statistics generated by our profiling tools (see: \autoref{sec:software-profiling-tools}).
We specifically look at the time spent on a set of runtime operations, for example, task creation and team barrier, and the number of tasks generated and executed by each thread.

\autoref{fig:imbalance} shows the time spent by each thread in various states for Fibonacci (\textbf{Fib}) and Multisort (\textbf{Sort}) applications.
In the left-hand figures, each stacked horizontal bar summarizes one of the 192 threads, 
with the colors representing, from left to right, tasking (green), task creation (blue), \emph{taskwait} directive (yellow), barrier (red), and stall (grey).

We focus on tasking (green), task creation (blue), \emph{taskwait} directive (yellow) and stall (grey) as barrier (red) is almost indistinguishable in the figure.
We consider utilized time as the time spent on tasking (green) and task creation (blue).
In the right chart, the horizontal bars represent tasks created (blue) and tasks executed per thread (green).
\textbf{Fib}'s Timeline Summary shows that threads with lower thread ID have much shorter utilized time than other threads; in the Task Count Summary, they generated and executed fewer tasks, indicating a clear imbalance in both utilization and task count per thread. We conclude that \textbf{Fib} is poorly balanced in terms of both utilization and task count per thread. For \textbf{Sort}, although the task count appears balanced, mid-range threads have higher utilization, leaving other threads idle and causing performance degradation.

\begin{figure*}[!ht]
\centering
    \includegraphics[width=\linewidth,trim=2mm 3mm 2mm 2mm,clip]{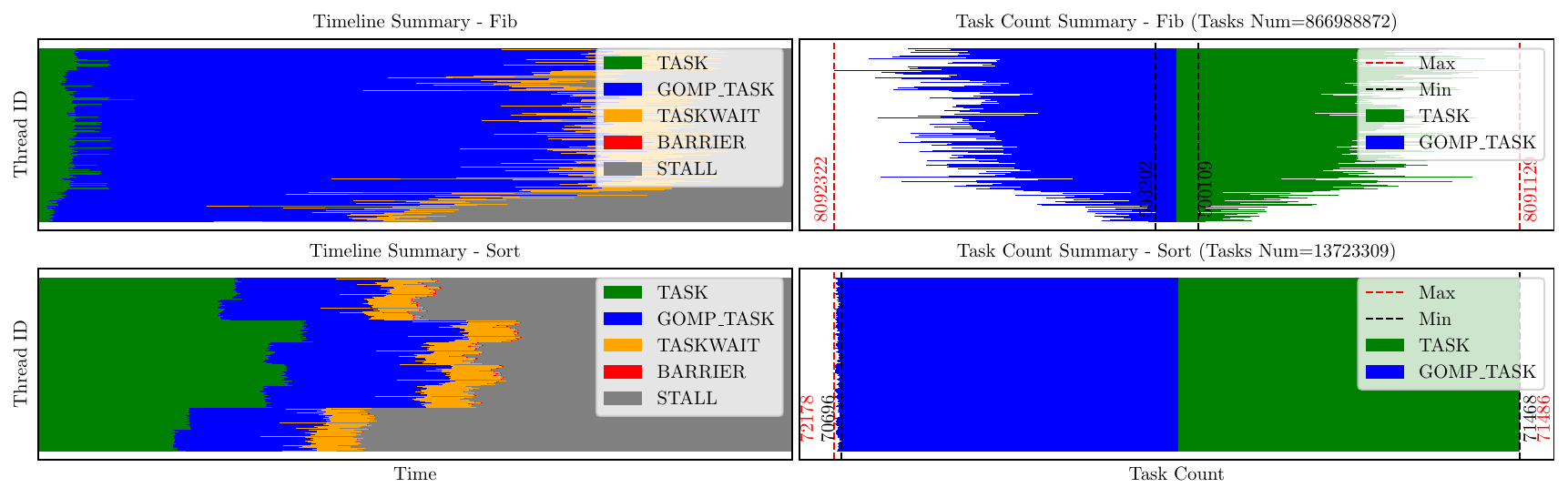}
    \caption{Load imbalance of \textbf{Fib} (above) and \textbf{Sort} (below) with XGOMP. 
    Y-axes are Thread ID.
    Left (Timeline Summary), X-axis is Time, showing the amount of time each thread spends in each state. Right (Task Count), blue bar shows the number of tasks generated, green bar shows the number of tasks executed.}
    \label{fig:imbalance}
\end{figure*}

\subsection{Lock-less Messaging Protocol}
\label{sec:dlb-commu}
Implementing lock-less DLB strategies first requires a lock-less messaging protocol for communicating between threads. We extend our prior work~\cite{NOOKALA2024444} with a conditionally random victim selection strategy~\cite{williams_libfork_2024}, as follows. 

We extend the OpenMP-defined thread data structure with two 64-bit memory cells: a \emph{round} cell and a \emph{request} cell.
The \emph{round} number is maintained by the victim to track steal requests that have been ``handled'' enabling potential thieves to check if the victim is able to be stolen from. It is a monotonically increasing number that is initialized to 1 and incremented by the victim each time a steal request is successfully handled. We consider successful handling of a steal request to be when the victim sees the request and processes whether the request is valid.
The \emph{request} cell contains a 40-bit round number and a 24-bit worker ID.
The thief sends a request to the victim by writing the victim's round number combined with the thief's worker ID to the \emph{request} cell.

% An idle worker becomes a thief. 
The thief messaging logic is in \autoref{algo:theif-logic}.
In all algorithms, we define $p_{\text{local}}$ as the probability that the thief steals within their local NUMA zone. $tid_{\text{i}}$ is the thread ID of worker $i$, while $ctid_{\text{thief}}$ is that of the current thief. $req$ and $round$ are the current request cell and round number, respectively. The number of tasks to steal per request is $N_{\text{steal}}$.
The thief picks a victim randomly, with probability $p_{\text{local}}$ of NUMA-local and $1-p_{\text{local}}$ of NUMA-remote.
The thief then compares the victim's \emph{round} cell and the round number extracted from the victim's \emph{request} cell.
If the round number from \emph{round} cell is greater than the round number from \emph{request} cell, meaning there has not been a new request for that victim, it then writes a request combining the victim's current round number from \emph{round} cell and the thief's worker ID to the the victim's \emph{request} cell; otherwise the request is not sent because there has been a request already.

The thief maintains a timeout counter to deal with long idle times.
This counter is incremented each time the thief reaches the same scheduling point while idle and is reset if 
i) the timeout counter reaches $T_{\text{interval}}$, triggering a retry, or ii) the worker is no longer idle.
Timeout can arise if a request is not sent or is overwritten by other thieves, or if a victim is idle.
When a worker finds a task to execute, it becomes a victim and tries to handle a request if one exists.
\begin{algorithm}[bth]
\begin{algorithmic}[1]
    \STATE $tid_{\text{victim}} \gets \texttt{pickVictimTID}(p_{local}, tid_{\text{i}})$
    \STATE $req \gets \texttt{getVictimRequest}(tid_{\text{victim}})$
    \STATE $round \gets \texttt{getVictimRound}(tid_{\text{victim}})$
    \STATE $curr \gets req \And ((1\ll40)-1)$ \COMMENT{extract round\phantom{X}}
    \IF{$curr < round$}
        \STATE $newReq \gets (tid_{\text{i}}<<40) \And round$
        \STATE $\texttt{setVictimRequest}(tid_{\text{victim}}, newReq)$
    \ENDIF
\end{algorithmic}

\vspace{1ex}

\caption{Thief worker send request logic}
\label{algo:theif-logic}
\end{algorithm}

We show the victim logic in \autoref{algo:victim-logic}.
A victim first checks its own \emph{request}. 
If the round number of its \emph{request} is equal to its current round number from its \emph{round} cell, this is considered a valid request. The victim then processes the request using a DLB strategy (NA-RP or NA-WS) and increments the round number so that it is willing to accept new steal requests.

\begin{algorithm}[bth]
\begin{algorithmic}[1]
     \STATE $tid_{\text{thief}} \gets req\gg40$
    \STATE $r_{\text{thief}} \gets req \And ((1\ll40)-1)$
    \IF{$r_{\text{thief}} == round$}
        \STATE $\texttt{doLoadBalancing}()$
        \STATE $round \gets round + 1$
        % \STATE $\texttt{setVictimRound}(round + 1)$
    \ENDIF
    
\end{algorithmic}

\vspace{1ex}

\caption{Victim worker handle request logic}
\label{algo:victim-logic}
\end{algorithm}

The lock-less inter-core communication overhead is primarily due to thieves accessing remote memory cells.
Unlike traditional inter-core communication, which often relies on atomic operations with typical lower-bound per-access latencies of around 100~ns~\cite{memlatency}, our approach can leverage multi-level, shared caches with lower-bound per-access latencies of just a few nanoseconds. 
This characteristic rewards spatially adjacent communications and task distribution.

\subsection{NUMA-aware Redirect Push (NA-RP)}
\label{sec:rp-design}

In our first DLB strategy, NUMA-aware Redirect Push \textbf{(NA-RP)}, the runtime redistributes tasks dynamically during work-sharing: see \autoref{algo:rp-logic}. 
The thief first repeats the logic in \autoref{algo:theif-logic} $N_{\text{victim}}$ times---in effect, asking $N_{\text{victim}}$ victims to redirect tasks to it. 
Each victim, upon successfully handling a request using \autoref{algo:victim-logic}, calls \texttt{doLoadBalancing()}: see \autoref{algo:rp-logic}.
\texttt{doLoadBalancing()} then changes victim state to be ready to redirect $N_{\text{steal}}$ new tasks to the thief: see \texttt{doRedirectPush()}.
If the thief's task queue is full or the victim handles all redirect push requests, the victim increments the round number and is ready to accept new request.

\begin{algorithm}[bth]
\begin{algorithmic}[1]
    \item[{\bf Function} \texttt{doLoadBalancing}():]{}
    \IF{$tid_{\text{thief}} == -1$}
        % \STATE $req \gets \texttt{getVictimRequest}(tid_{\text{i}})$
        \STATE $ctid_{\text{thief}} \gets req\gg40$
        \STATE $pushed\_tasks \gets 0$
    \ENDIF
\vspace{1ex}
    \item[{\bf Function} \texttt{doRedirectPush}($newTask$):]{}
        \IF{$pushed\_tasks \geq N_{\text{steal}} \OR \texttt{isTargetQFull}(ctid_{\text{thief}},tid_{\text{i}})$}
            \STATE $ctid_{\text{thief}} \gets -1$ \COMMENT{No thief\phantom{xx}}
        \ELSE
            \STATE $\texttt{pushXQueueTask}(ctid_{\text{thief}}, newTask)$
            \STATE $pushed\_tasks \gets pushed\_tasks + 1$
        \ENDIF
\end{algorithmic}

\vspace{1ex}

\caption{Redirect push logic}
\label{algo:rp-logic}
\end{algorithm}

% Performance Implication
The NA-RP strategy extends XQueue's lock-less enqueuing mechanisms by only altering the target task queue to which new tasks are pushed. 
It prioritizes helping underloaded workers to mitigate work imbalance.
The additional overhead lies in the exchange of messages across cores.

\subsection{NUMA-aware Work Stealing (NA-WS)}
\label{sec:ws-design}

In our second DLB strategy, NUMA-aware Work Stealing \textbf{(NA-WS)},  %, %instead of redirecting the newly created tasks to the thief,
the thief steals tasks from specific victims based on their relative NUMA location. 

We first explore a queue-based strategy. 
Recall that each XQueue thread possesses $N_{\text{worker}} \times S_{\text{queue}}$ task queues where $N_{\text{worker}}$ is the number of workers and $S_{\text{queue}}$ is the size of each individual queue. 
We allocate $N_{\text{worker}}$ \emph{request} cells and \emph{round} cells that map to each queue.
The routine starts with the thief picking victims with $p_{\text{local}}$ of picking NUMA-local workers.
It then randomly picks a queue in each of these victims, sending requests using our lock-less communication protocol.
Upon receiving requests, a potential victim scans a subset at a time to find one to further process. 
This strategy guarantees that there is only one producer and consumer for each request cell, thus avoiding requests being overwritten, which could result in retry delays for the thieves. 
Through experiments, we observed that this method does not mitigate work imbalance and it introduces additional overhead.
With millions of steal requests sent, only few tasks are stolen. 
We observed that 62\% of requests are handled by victims. 
However, less than 1\% of all requests found by victims are valid, and around 0.01\% of these valid requests result in successful steals. 
The runtime is rarely able to steal because there is a mismatch between the number of requests sent and the number of requests handled. 

We further improve this strategy by letting a thief steal a batch of tasks from a victim, rather than requesting from a victim's individual task queues: thus reducing communication from queue-to-queue to worker-to-worker. 
The thief uses the logic in \autoref{algo:theif-logic}.
A victim, upon successfully handling a request,
dequeues up to $N_{\text{steal}}$ tasks from its queue and enqueues them to the thief's target task queue. This round of stealing completes when 1) no task is found from the victim; 2) the thief's target task queue is full; or 3) $N_{\text{steal}}$ tasks have been migrated. We present the pseudo-code in \autoref{algo:ws-logic}.

\begin{algorithm}[bth]
\begin{algorithmic}[1]
    \item[{\bf Function} doLoadBalancing():]{}
    \STATE $tid_{\text{thief}} \gets req\gg40$
    \STATE $pushed\_tasks \gets 0$
    \WHILE{$ (!\texttt{isTargetQFull}(tid_{\text{thief}}, tid_{\text{i}})) \AND (!\texttt{isMyQEmpty}())$}
        \STATE $task \gets \texttt{removeXQueueTask}(tid_{\text{i}})$
        \STATE $\texttt{pushXQueueTask}(tid_{\text{thief}}, task)$
        \STATE $pushed\_tasks \gets pushed\_tasks + 1$
        \IF{$pushed\_tasks \geq N_{\text{steal}}$}
            \STATE break
        \ENDIF
    \ENDWHILE
\end{algorithmic}

\vspace{1ex}

\caption{Work stealing design}
\label{algo:ws-logic}
\end{algorithm}

% Performance implication
Our lock-less work stealing approach inherits both lock-less enqueuing and dequeuing mechanisms from XQueue by migrating already-queued tasks from one worker's queue to another instead of co-locating the newly spawned tasks. 
Theoretically, this strategy incurs slightly more overhead compared to NA-RP because of the additional dequeuing operations. However, it could result in far fewer tasks stolen than tasks redirected in the NA-RP approach because stealing $N_{\text{steal}}$ tasks is the upper bound for each round of stealing, while redirecting $N_{\text{steal}}$ is the upper and lower bound for each round of redirection. 
Unlike NA-RP, which tends to push tasks away from where they are created to remote workers, NA-WS tends to bring the tasks back to where they are created, thus exploiting benefits of data proximity.

\subsection{DLB Configuration}
\label{subsec:dlb-config}
We implement several configurable parameters to control 
how the DLB strategies operate. These parameters are
the number of victims ($N_{\text{victim}}$) to whom a worker sends requests each time it becomes a thief; 
the max number of tasks to be stolen/redirected ($N_{\text{steal}}$) for each request; 
the timeout interval ($T_{\text{interval}}$) between two requests (addressing the situation that an underloaded worker could remain idle for an extended period of time if requests sent while idle are not answered); and NUMA-local probability ($P_{\text{local}}$) controlling the probability that a thief steals from NUMA-local nodes.

\section{Software Profiling Tools}
\label{sec:software-profiling-tools}

To better analyze the performance characteristics of OpenMP tasking we developed new per-thread performance profiling tools.
These tools enable us to track the time spent on tasking runtime operations.
We categorize the operations that are most important for performance analysis into several event types.
We identify the start and end of each event and insert marker functions \texttt{perf\_record} with corresponding settings to record the event to profiler memory. 
We use \emph{rdtscp} (ReaD Time-Stamp Counter and Processor) id value as the timestamp for each event. 
\emph{rdtscp} is a light-weight processor intrinsic instruction that reads the current value of the processor's timestamp counter.
The processor increments the timestamp counter monotonically every clock cycle and resets it to 0 when the processor is reset.
Note that while \emph{rdtscp} is not a serializing instruction, it ensures that all prior instructions have executed and all previous loads are globally visible~\cite{rdtscp}. 
At the end of the program, the \texttt{xomp\_perflog\_dump} API can be used to save the logs to the file system in a path defined by environment variables.

Although enabling performance logging introduces overheads for fine-grained tasks, for example, due to increased memory accesses when storing event information and the unavoidable hardware overhead of \emph{rdtscp}, it still captures overall runtime behaviors, as each logging operation overhead is almost constant. 
Below, we describe our event classes. Each event class includes a pair of start and end timestamps.

The cycles spent by a thread for different purposes are tracked as follows:
By a task: \texttt{TASK};
creating tasks (crucial because fine-grained tasks can spend a large portion of their lifecycle on task creation): \texttt{GOMP\_TASK};
and waiting: \texttt{TASKWAIT}.
\texttt{BARRIER} denotes the number of cycles spent on a barrier. If a thread is unoccupied, i.e., no task is scheduled in the current thread and it is busy checking its task queue, we record the cycles as \texttt{STALL}.

We also instrument per-thread statistical counters. These counters incur little overhead as they are thread-local and can use lower-level caches.
We implement a set of general performance counters and specialized counters for our DLB strategies. 
To track the locality of a task, we assign the thread ID to the task upon creation.

To manage data locality, XGOMPTB tracks the number of tasks executed: i) by the thread that created it (\texttt{NTASKS\_SELF}); ii) by the NUMA node that created it (\texttt{NTASKS\_LOCAL}); and iii) by another node (\texttt{NTASKS\_REMOTE}).
XGOMPTB also tracks the number of tasks that are i) statically pushed (\texttt{NTASKS\_STATIC\_PUSH} and ii) are not pushed due to the target queue being full and are instead executed immediately (\texttt{NTASKS\_IMM\_EXEC}). 

In addition, for the two DLB strategies we track the number of steal requests (\texttt{NREQ\_SENT}), handled requests (\texttt{NREQ\_HANDLED}), and stolen tasks (\texttt{NTASKS\_STOLEN}).
Handled requests are further categorized into requests that i) result in a task being stolen (\texttt{NREQ\_HAS\_STEAL}); or ii) fail due to an attempt to steal tasks from an empty queue (\texttt{NREQ\_SRC\_EMPTY}) or to a full queue (\texttt{NREQ\_TARGET\_FULL}).
Stolen tasks are further categorized into tasks stolen by i) NUMA-local thieves or ii) NUMA-remote thieves.

% EVALUATION

\section{Evaluation}
\label{sec:eval}
% Num Tasks:
% Fib(40(DLB)): 331,160,280 - 331M
% Fib(42): 866,988,872 - 866M
% size(fib) = 10 - 100; dom: 10
% NQueens(15): 2,532,748,320 - 2.5B
% size(nqueens) = 100 - 1000; dom: 100
% FFT(268M): 18,280,176 - 18M
% FFT(536M): 34,533,104 - 34M
% size(FFT) = 1000 - 1000,000: 1e3 - 1e6; dom:1e3,4,5,6
% Floorplan(15): 19,043,928 - 19M 
% Floorplan(20): 66,890,223 - 66M
% size(floorplan) = 1000 - 1000,000: 1e3 - 1e6; dom: 1e2,3; 4,5,6
% Health(large): 126,363,001 - 126M
% size(health) = 1000 - 100,000: 1e3 - 1e5; dom: 1e3;4,5
% UTS(tiny): 30,399,117 - 30M
% UTS(small): 111,345,631 - 111M
% size(UTS) = 1000-1000,000: dom: 1e3,4;5,2,6
% STRAS(2048): 960800 - 960K
% STRAS(8192): 47,079,208 - 47M
% size(STRAS) = 1e4-9; dom: 1e4,6,7,5,8,9
% Sort(268M): 3,075,031 - 3M
% Sort(1B): 13,723,309 - 13M
% size(sort) = 1e3-1e5; dom: 1e5, 3, 6, 4
% Alignment(1000):  499,500 - 499K
% size(alignment): 1e5-7; dom: 1e6, 7, 5
We evaluate our XGOMP and XGOMPTB implementations and NA-RP and NA-WS strategies using nine applications from the Barcelona OpenMP Task Suite (BOTS)~\cite{duran_barcelona_2009}: 
Fibonacci (\textbf{Fib}), \textbf{NQueens}, Fast Fourier Transform (\textbf{FFT}),
Floor Plan (\textbf{FP}), \textbf{Health}, Unbalance Tree Search (\textbf{UTS}), Strassen Matrix Multiplication (\textbf{STRAS}), and Protein Alignment (\textbf{Align}).
We first compare XGOMP with the original GNU OpenMP (GOMP). 
We also include two LLVM variants---LLVM OpenMP (LOMP) and LLVM OpenMP using XQueue (XLOMP)---in our results for comparison.
We conduct all experiments on an Intel Skylake-192 machine, with 192 cores (384 hardware threads) and eight NUMA zones.
We compile the benchmarks with GNU GCC version 12.2.1 and LLVM version 14.0.0, both with the \emph{-fopenmp} flag and O3 optimization level.
We use up to 192 threads and bind OpenMP threads to cores with \emph{close} thread affinity.
We use the following input arguments for our applications.
\textbf{Fib}: 42, \textbf{NQueens}: 15, \textbf{FFT}: 536M, \textbf{FP}: 20, \textbf{Health}: large, \textbf{UTS}: small, \textbf{STRAS}: 8192 (Y=16), \textbf{Sort}: 1B, and \textbf{Align}: 1000.
For our DLB experiments (\autoref{sec:eval-dlb}), in which we run large parameter sweeps over many configuration parameters, we scale down \textbf{Fib}: 40, \textbf{FFT}: 268M and \textbf{Sort}: 268M, \textbf{FP}: 15, \textbf{UTS}: tiny, and \textbf{STRAS}: 2048 (Y=16) to reduce experiment times. 
We have validated experimentally that the larger and smaller experiments yield similar results.

\subsection{GNU OpenMP with XQueue and Distributed Tree Barrier}
\label{sec:eval-static}

\autoref{fig:xgomp-exec-time} shows the average application execution time of the nine benchmarks with different OpenMP implementations.
Each bar shows the average execution time of five 192-thread runs. 
We use our profiling tools to measure task size (in \emph{rdtscp} cycles) and order applications based on their task size (from small to large).
We see across all benchmarks, our XQueue-based implementations (XGOMP, XGOMPTB, and XLOMP) as well as LOMP are several orders of magnitude faster than GOMP.

To understand these results, we measure the time spent on various runtime profiling events (see \autoref{sec:software-profiling-tools}).
We observe that applications for which LOMP and XLOMP outperform GOMP and XGOMP---\textbf{Fib}, \textbf{NQueens}, \textbf{FP}, \textbf{Health}, and \textbf{UTS}---spend much of thread time on task creation (mostly task allocation).
While applications for which XGOMP and XGOMPTB outperform XLOMP---\textbf{FFT}, \textbf{STRAS}, \textbf{Sort}, and \textbf{Align}---spend most of their time on task execution.
These results are due to the way that memory is allocated. All OpenMP implementations use \emph{malloc} to allocate memory for tasks. \emph{malloc} is thread safe and thus uses synchronization. The default allocation approach used by Linux prioritizes allocating memory in the NUMA node closest to the CPU executing the task.
GOMP invokes \emph{malloc} each time it creates a new task, whereas LOMP instead allocates memory using a fast multi-level allocator, which pre-allocates chunks of buffer space and manages this space itself. When a thread creates a new task, LOMP/XLOMP will i) use a local buffer if available; ii) synchronously but locality-agnostically acquire a buffer from another thread if available; or iii) call \emph{malloc}.
For fine-grained tasks, large portions of time are spent on task allocation/de-allocation. For GOMP, threads contend for \emph{malloc} operations, which leads to sequential \emph{malloc}s. LOMP can parallelize memory allocation, because method i) is frequently used.
For large tasks, LOMP/XLOMP performs a similar number of \emph{malloc} requests to GOMP as the multi-level allocator is less beneficial. That is, methods i) and ii) are less likely to be successful (as longer tasks consume the buffer for longer periods). Further, LOMP “steals” buffer space using method ii), which results in more tasks executed remotely. LOMP has slightly higher task execution overhead because it handles a richer set of cases.

\textbf{FFT} task sizes range between 10\textsuperscript{2}--10\textsuperscript{6} cycles, with the highest proportion around 10\textsuperscript{3}--10\textsuperscript{4} cycles;
\textbf{STRAS} task sizes range between 10\textsuperscript{3}--10\textsuperscript{7} cycles with most around 10\textsuperscript{4} cycles.
\textbf{Sort} is similar to \textbf{STRAS}, with most sizes around 10\textsuperscript{5} cycles.
\textbf{Align} task sizes are between 10\textsuperscript{5}--10\textsuperscript{7} cycles, with the highest proportion around 10\textsuperscript{6} cycles.
These task sizes allow XGOMP and XGOMPTB to outperform LOMP and XLOMP  
as the benefit of the multi-level allocator is reduced.
\textbf{Health} has average task size greater than \textbf{FFT}; however, it has many more tasks concentrated at lower sizes(10\textsuperscript{3}--10\textsuperscript{4} cycles) which benefit from the multi-level allocator.

% XGOMP EXEC TIME
\begin{figure}[!ht]
    \centering
    \includegraphics[width=\columnwidth,trim=2mm 3mm 2mm 2.5mm,clip]{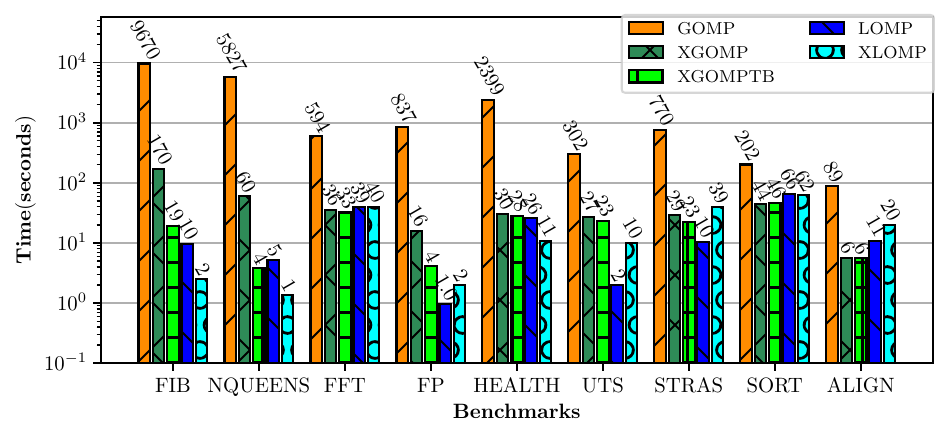}
    \caption{Absolute execution time of BOTS benchmarks (lower is better).}
    \label{fig:xgomp-exec-time}
\end{figure}

\begin{figure}[!ht]
    \centering
    \includegraphics[width=\columnwidth,trim=2mm 3mm 2mm 2.5mm,clip]{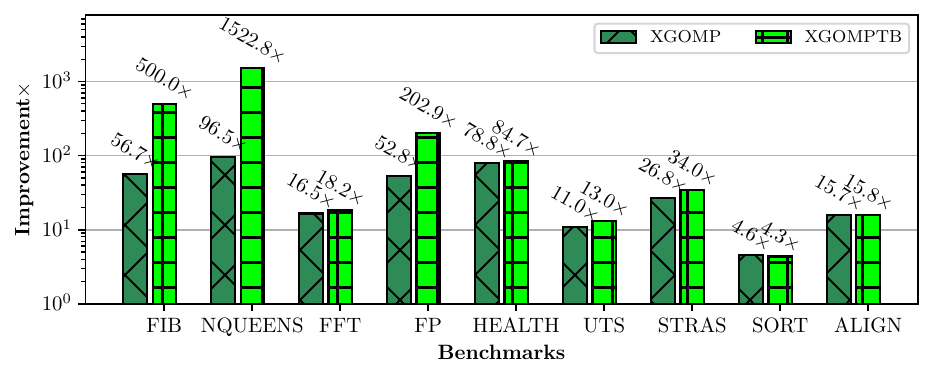}
    \caption{XGOMP/XGOMPTB performance improvement over GOMP, 192 threads (higher is better).}
    \label{fig:xgomp-improve}
\end{figure}

\begin{figure*}[!ht]
    \centering
    \includegraphics[width=\textwidth,trim=2.5mm 3mm 2mm 2mm,clip]{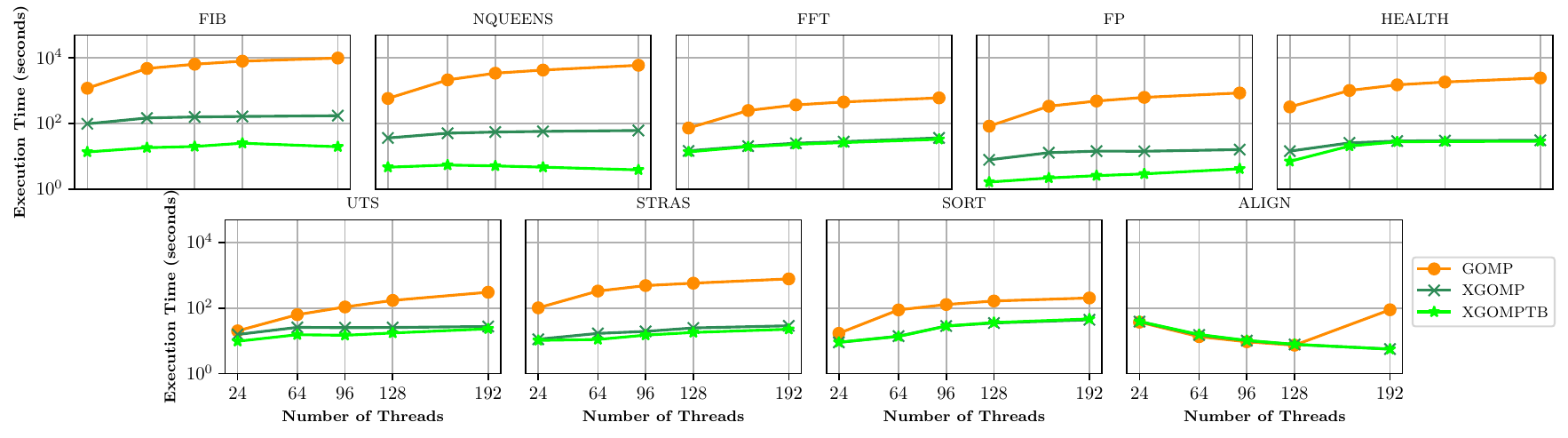}
    \caption{Scaling performance of different methods comparing execution time as the number of threads is increased for each BOTS application (lower is better)} 
    \label{fig:xgomp-thr-scaling}
\end{figure*}

In \autoref{fig:xgomp-improve} we show the performance improvement of XGOMP and XGOMPTB over GOMP. 
We see that our methods enable performance improvements of up to 96.5$\times$ for XGOMP and 1522.8$\times$ for XGOMPTB when compared to GOMP.
All benchmark applications benefit from XQueue. Applications with smaller task sizes benefit more from the distributed tree barrier (e.g. Fib, NQueens, FP); while those with larger tasks benefit least from the barrier.

% XGOMP performance with thread scaling
\autoref{fig:xgomp-thr-scaling} shows the performance of GOMP, XGOMP and XGOMPTB with increasing number of threads for each application.
XGOMP and XGOMPTB perform significantly better than GOMP for most applications and thread counts.
For \textbf{Align}, which has the largest average task size, we see comparable performance at lower thread counts due to extremely low lock contention.
For all applications, performance improvement grows as we increase from 24 threads (one socket) to 192 threads (eight sockets). 
However, application performance does not scale linearly with threads due to work inflation~\cite{olivier2013characterizing}, that is, due to increased costs for operations performed in a parallel vs.\ a single-processor run.
Work inflation can be due to many factors, such as work migration, shared use of an LLC, and the need to access data on remote sockets; it can be reduced by placing computations and data on the same socket, removing the need for remote memory access~\cite{deters_numa-aware_2018}.

\subsection{XGOMPTB with Dynamic Load Balancing}
\label{sec:eval-dlb}

We study our two lock-less DLBs with the same BOTS benchmarks using the Skylake-192 machine. 
We conduct experiments with 192 threads on eight NUMA zones, controlling runtime behavior with the configuration parameters described in \autoref{subsec:dlb-config}.
For each application, we conduct a parameter sweep to identify the optimal settings for each value of $N_{\text{victim}}$, $N_{\text{steal}}$, $T_{\text{interval}}$ and $P_{\text{local}}$.
We run each DLB configuration ten times and compare its average execution time with XGOMPTB. 

\begin{figure}[!ht]
    \centering
    \includegraphics[width=\columnwidth,trim=2.5mm 3mm 2mm 2mm,clip]{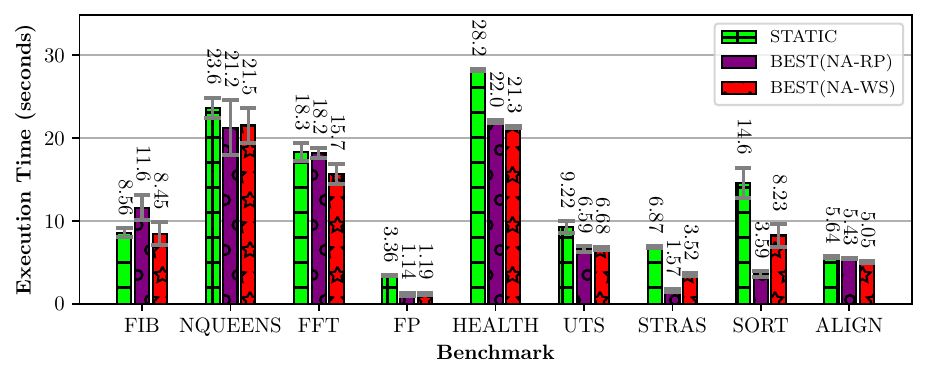}
    \caption{XGOMP DLB performance comparisons on BOTS with best performance settings, 192 threads.}
    \label{fig:best-all}
\end{figure}

\begingroup
\addtolength{\tabcolsep}{-5.5pt}
\renewcommand{\arraystretch}{1}
\begin{table}[!ht]
    \centering
    \caption{Optimal DLB Settings for Redirect Push and Work Stealing}
    \resizebox{\columnwidth}{!}{%
    \begin{tabular}{|c|rr|rr|rr|rr|rr|rr|rr|rr|rr|}
    \hline
    \multirow{1}{*}{Benchmark} & \multicolumn{2}{c|}{\textbf{Fib}} & \multicolumn{2}{c|}{\textbf{NQueens}} & \multicolumn{2}{c|}{\textbf{FFT}} & \multicolumn{2}{c|}{\textbf{FP}} & \multicolumn{2}{c|}{\textbf{Health}} & \multicolumn{2}{c|}{\textbf{UTS}} & \multicolumn{2}{c|}{\textbf{STRAS}} & \multicolumn{2}{c|}{\textbf{Sort}} & \multicolumn{2}{c|}{\textbf{Align}} \\
    \cline{1-19}
    DLB strat. & RP & WS & RP & WS & RP & WS & RP & WS & RP & WS & RP & WS & RP & WS & RP & WS & RP & WS\\

    \hline
    $N_{\text{victims}}$ & 1 & 1 & 24 & 16 & 24 & 24 & 24 & 16 & 24 & 8 & 8 & 8 & 24 & 8 & 24 & 24 & 8 & 16\\
    % \hline
    $N_{\text{steals}}$ & 16 & 1 & 16 & 1& 1 & 32 & 32 & 32 & 32 & 32 & 1 & 32 & 32 & 32 & 32 & 32 & 8 & 8 \\
    % \hline
    $T_{\text{interval}}$ & 10\textsuperscript{5} & 10\textsuperscript{4} & 10\textsuperscript{5} & 10\textsuperscript{4} & 10\textsuperscript{3} & 10\textsuperscript{4} & 10\textsuperscript{5} & 10\textsuperscript{5} & 10\textsuperscript{3} & 10\textsuperscript{3} & 10\textsuperscript{4} & 10\textsuperscript{5} & 10\textsuperscript{4} & 10\textsuperscript{3} & 10\textsuperscript{3} & 10\textsuperscript{3} & 10\textsuperscript{4} & 10\textsuperscript{4}\\
    % \hline
    $P_{\text{local}}$ & 1 & 1 & 1 & 1 & 1 & 1 & 1 & 1 & 1 & 0.5 & 1 & 1 & 1 & 0.03 & 1 & 0.03 & 0.03 & 1 \\
    \hline
    \end{tabular}
    }
    \label{tab:combined-settings}
\end{table}
\endgroup %tab:combined-settings

\autoref{fig:best-all} compares the best average performance of each DLB strategy (NA-RP and NA-WS) with XGOMPTB which uses static load balancing (SLB).
The settings used to obtain the best performance for NA-RP and NA-WS are presented in \autoref{tab:combined-settings}.
All benchmarks except \textbf{Fib} show performance improvement.
We re-run the experiments, collecting statistics for DLB in \autoref{tab:bots-stats} and SLB in \autoref{tab:bots-stats-static}.
To understand the effect of locality, we present the number of tasks executed along with their origin: tasks created on the same core (self), tasks created by a worker within the same NUMA zone (local), and tasks created by a worker in a different NUMA zone (remote). Tasks that run on the same core they were created on will generally have more first-level cache hits than those that do not. Tasks that run in the same NUMA zone where they were created will have fewer first-level cache hits, but will still have more shared cache hits and faster memory access than those that run in a remote NUMA zone.
Tasks that are executed immediately, will do so on the same core (self) and thus benefit from the low-level cache.
We show the number of tasks pushed with SLB and tasks executed immediately due to a failed push.
The number of requests sent, handled, and that result in tasks stolen demonstrate the efficacy of our lock-less messaging protocol.

Finally, we look at the locality of stolen tasks---more locally stolen tasks means the system can benefit from faster memory access compared to remote memory access.
Due to profiling overhead and randomness, the results presented are slightly different from those in \autoref{fig:best-all}.

\subsubsection{NUMA-aware Redirect Push (NA-RP)}
\label{sec:eval-narp}

% Fib
We see in \autoref{fig:best-all} that \textbf{Fib} performance degrades with NA-RP.
This is because NA-RP pushes more tasks away, incurring a cost of more than 100~ns for each task that could otherwise be self-executed within nanoseconds.
It has the largest $T_{\text{interval}}$, the smallest $N_{\text{victim}}$, and a moderate $N_{\text{steal}}$.
We see in \autoref{tab:bots-stats} that NA-RP results in 8.2M stolen tasks and the fewest self-executed tasks across all strategies.
\textbf{Fib}'s tasks are small, 10--80 cycles, and its DAG has a long critical path that can barely benefit from parallelism.
Many tasks are redirected, removing the potential locality benefit provided by the first-level cache.
Most steal requests are successfully handled, which demonstrates the efficacy of our lock-less messaging protocol for fine-grained tasks.

\begingroup
\addtolength{\tabcolsep}{-5pt}
\renewcommand{\arraystretch}{0.95}
\begin{table*}[!ht]
\scriptsize
\caption{BOTS Runtime Statistics with NA-RP and NA-WS DLB strategies. Reporting average from 10 experiments. \label{tab:bots-stats}}
\resizebox{\textwidth}{!}{%
\begin{tabular}{lrr|rr|rr|rr|rr|rr|rr|rr|rr|}
\hline
\multicolumn{1}{|l|}{Benchmark} & \multicolumn{2}{c|}{\textbf{Fib}} & \multicolumn{2}{c|}{\textbf{NQueens}} & \multicolumn{2}{c|}{\textbf{FFT}}  & \multicolumn{2}{c|}{\textbf{FP}} & \multicolumn{2}{c|}{\textbf{Health}} & \multicolumn{2}{c|}{\textbf{UTS}} & \multicolumn{2}{c|}{\textbf{STRAS}} & \multicolumn{2}{c|}{\textbf{Sort}} & \multicolumn{2}{c|}{\textbf{Align}} \\
\hline
\multicolumn{1}{|l|}{DLB strat.} & RP & WS & RP & WS & RP & WS & RP & WS & RP & WS & RP & WS & RP & WS & RP & WS & RP & WS \\
\hline
\multicolumn{1}{|l|}{Time (secs)} & 9.9 & 8.0 & 26.5 & 23.2 & 17.0 & 15.5 & 1.3 & 1.3 & 21.2 & 20.9 & 6.4 & 6.6 & 1.6 & 3.7 & 3.8 & 9.5 & 5.6 & 5.2 \\
\multicolumn{1}{|l|}{Self tasks} & 310.8 M & 318.2 M & 2.4 B & 2.5 B & 946.7 K & 1.7 M & 17.1 M & 16.5 M & 33.8 M & 33.5 M & 6.2 M & 347.1 K & 277.2 K & 416.6 K & 146.3 K & 63.5 K & 2.9 K & 2.6 K \\
\multicolumn{1}{|l|}{Local tasks} & 12.2 M & 1.6 M & 30.1 M & 6.5 M & 11.6 M & 2.0 M & 1.2 M & 302.3 K & 41.2 M & 10.8 M & 20.4 M & 3.6 M & 553.8 K & 65.7 K & 2.7 M & 376.0 K & 59.8 K & 59.8 K \\
\multicolumn{1}{|l|}{Remote tasks} & 8.2 M & 11.4 M & 74.1 M & 47.4 M & 5.7 M & 14.6 M & 837.6 K & 2.3 M & 51.3 M & 82.1 M & 3.9 M & 26.5 M & 129.7 K & 478.5 K & 208.9 K & 2.6 M & 436.8 K & 437.0 K \\
\multicolumn{1}{|l|}{Static push} & 12.3 M & 13.0 M & 90.9 M & 54.2 M & 15.1 M & 16.7 M & 1.1 M & 2.6 M & 64.1 M & 93.5 M & 20.6 M & 30.3 M & 233.0 K & 548.8 K & 484.5 K & 3.0 M & 496.6 K & 496.9 K \\
\multicolumn{1}{|l|}{Imm. exec} & 310.6 M & 318.1 M & 2.4 B & 2.5 B & 468.4 K & 1.6 M & 17.0 M & 16.5 M & 33.3 M & 32.9 M & 5.4 M & 123.2 K & 272.7 K & 412.0 K & 134.0 K & 41.9 K & 2.9 K & 2.6 K \\
\multicolumn{1}{|l|}{Req. sent} & 590.1 K & 1.6 M & 1.3 M & 2.4 M & 4.2 M & 16.3 M & 31.3 K & 884.1 K & 976.8 K & 13.2 M & 5.5 M & 19.1 M & 18.9 K & 233.5 K & 86.8 K & 5.2 M & 1.8 K & 266.4 K \\
\multicolumn{1}{|l|}{Req. handled} & 514.9 K & 1.3 M & 872.9 K & 2.3 M & 2.7 M & 10.9 M & 28.5 K & 798.0 K & 905.9 K & 12.0 M & 4.4 M & 12.5 M & 14.3 K & 171.6 K & 76.9 K & 2.0 M & 192  & 265.8 K \\
\multicolumn{1}{|l|}{Req. w/ steal} & 514.7 K & 576.4 K & 872.7 K & 1.3 M & 2.7 M & 2.4 M & 28.3 K & 395.9 K & 904.7 K & 7.9 M & 4.4 M & 4.8 M & 14.1 K & 114.7 K & 76.7 K & 683.2 K & 0  & 156.4 K \\
\multicolumn{1}{|l|}{Total steal} & 8.2 M & 576.4 K & 14.0 M & 1.3 M & 2.7 M & 19.5 M & 909.1 K & 7.0 M & 29.0 M & 176.9 M & 4.4 M & 48.9 M & 455.1 K & 2.8 M & 2.5 M & 9.0 M & 0  & 861.0 K \\
\multicolumn{1}{|l|}{Local steal} & 8.2 M & 576.4 K & 14.0 M & 1.3 M & 2.7 M & 19.5 M & 909.1 K & 7.0 M & 29.0 M & 98.9 M & 4.4 M & 48.9 M & 455.1 K & 403.8 K & 2.5 M & 1.7 M & 0  & 861.0 K \\
\hline

\end{tabular}
}
\end{table*}
\endgroup

 % tab:bots-stats
\begingroup
\addtolength{\tabcolsep}{-5.5pt}
\renewcommand{\arraystretch}{1}
\begin{table}[!ht]
\caption{BOTS Runtime Statistics using SLB.}
\label{tab:bots-stats-static}
\centering
\resizebox{\columnwidth}{!}{%
\begin{tabular}{c|c|c|c|c|c|c|c|c|c}
\hline
\multicolumn{1}{|l|}{Benchmark} & \multicolumn{1}{c|}{\textbf{Fib}} & \multicolumn{1}{c|}{\textbf{NQueens}} & \multicolumn{1}{c|}{\textbf{FFT}} & \multicolumn{1}{c|}{\textbf{FP}} & \multicolumn{1}{c|}{\textbf{Health}} & \multicolumn{1}{c|}{\textbf{UTS}} & \multicolumn{1}{c|}{\textbf{STRAS}} & \multicolumn{1}{c|}{\textbf{Sort}} & \multicolumn{1}{c|}{\textbf{Align}} \\
\hline
\multicolumn{1}{|l|}{Time (secs)} & \multicolumn{1}{r|}{8.1} & \multicolumn{1}{r|}{23.8} & \multicolumn{1}{r|}{19.1} & \multicolumn{1}{r|}{3.4} & \multicolumn{1}{r|}{28.2} & \multicolumn{1}{r|}{9.0} & \multicolumn{1}{r|}{6.7} & \multicolumn{1}{r|}{11.1} & \multicolumn{1}{r|}{5.7} \\
\multicolumn{1}{|l|}{Self tasks} & \multicolumn{1}{r|}{318.1 M} & \multicolumn{1}{r|}{2.5 B} & \multicolumn{1}{r|}{203.0 K} & \multicolumn{1}{r|}{16.7 M} & \multicolumn{1}{r|}{658.2 K} & \multicolumn{1}{r|}{953.4 K} & \multicolumn{1}{r|}{5.1 K} & \multicolumn{1}{r|}{16.1 K} & \multicolumn{1}{r|}{3.0 K} \\
\multicolumn{1}{|l|}{Local tasks} & \multicolumn{1}{r|}{1.6 M} & \multicolumn{1}{r|}{6.1 M} & \multicolumn{1}{r|}{2.2 M} & \multicolumn{1}{r|}{287.9 K} & \multicolumn{1}{r|}{15.1 M} & \multicolumn{1}{r|}{3.5 M} & \multicolumn{1}{r|}{115.1 K} & \multicolumn{1}{r|}{368.3 K} & \multicolumn{1}{r|}{59.8 K} \\
\multicolumn{1}{|l|}{Remote tasks} & \multicolumn{1}{r|}{11.5 M} & \multicolumn{1}{r|}{44.2 M} & \multicolumn{1}{r|}{15.9 M} & \multicolumn{1}{r|}{2.1 M} & \multicolumn{1}{r|}{110.6 M} & \multicolumn{1}{r|}{25.9 M} & \multicolumn{1}{r|}{840.6 K} & \multicolumn{1}{r|}{2.7 M} & \multicolumn{1}{r|}{436.7 K} \\
\multicolumn{1}{|l|}{Static push} & \multicolumn{1}{r|}{13.2 M} & \multicolumn{1}{r|}{50.6 M} & \multicolumn{1}{r|}{18.2 M} & \multicolumn{1}{r|}{2.4 M} & \multicolumn{1}{r|}{126.4 M} & \multicolumn{1}{r|}{29.6 M} & \multicolumn{1}{r|}{960.8 K} & \multicolumn{1}{r|}{3.1 M} & \multicolumn{1}{r|}{496.5 K} \\
\multicolumn{1}{|l|}{Imm. exec} & \multicolumn{1}{r|}{318.0 M} & \multicolumn{1}{r|}{2.5 B} & \multicolumn{1}{r|}{108.2 K} & \multicolumn{1}{r|}{16.7 M} & \multicolumn{1}{r|}{0 } & \multicolumn{1}{r|}{799.2 K} & \multicolumn{1}{r|}{0 } & \multicolumn{1}{r|}{0 } & \multicolumn{1}{r|}{3.0 K} \\

\hline
\end{tabular}
}
\end{table}
\endgroup

% NQueen
For \textbf{NQueens}, NA-RP is the best performing strategy.
However, it has the largest error bar in \autoref{fig:best-all} indicating perhaps sensitivity to when steals occur.
We see in \autoref{tab:bots-stats} that NA-RP yields the worst average performance. 
We suspect that the profiling overhead may affect the dispatching pattern with SLB, which potentially results in more tasks being pushed to a remote node. 
\textbf{NQueens} works best when it is completely NUMA-local and the $T_{\text{interval}}$ is also the largest, while $N_{\text{victim}}$ and $N_{\text{steal}}$ are large, indicating that it benefits from large batches of steal requests after a long period of time.
\textbf{NQueens} with NA-RP yields the worst performance with profiling on because those runs have the least self-executed tasks, thus losing some of the benefit provided by the first-level cache. 
NA-RP with profiling also distributes more tasks to NUMA-remote nodes, which also increases the latency.

% FFT
\textbf{FFT} performs slightly better with NA-RP than with SLB.
It works best with the largest $N_{\text{victim}}$, and smallest $N_{\text{steal}}$ and $T_{\text{interval}}$.
It also uses fully NUMA-local.
Statistics show that NA-RP has the most local tasks, 4.7$\times$ more same-core tasks than SLB, and the fewest remote tasks.

\textbf{FP} performs 2.6$\times$ better with NA-RP than with SLB.
It works best with the largest $N_{\text{victim}}$, $N_{\text{steal}}$, $T_{\text{interval}}$, and $P_{\text{local}}$.
\textbf{FP} task sizes are varied, with the highest proportion around 10\textsuperscript{2}--10\textsuperscript{3} cycles and many ranging between 10\textsuperscript{3}--10\textsuperscript{6} cycles.
Scheduling tasks with large and varied task sizes could result in load imbalance.
NA-RP mitigates imbalance by redirecting 909K tasks.
1.3M more tasks are brought to the same core and NUMA-local than with SLB.
Most of its steal requests are successful.

\textbf{Health} performs 28.2\% better with NA-RP than with SLB.
It works best with the largest $N_{\text{victim}}$, $N_{\text{steal}}$, and $P_{\text{local}}$, and the smallest $T_{\text{interval}}$.
NA-RP brings 33.1M more tasks to the same core and 33.3M more tasks to be immediately executed which benefits from the low-level cache.
Most of its steal requests are successful.

\textbf{UTS} performs 40.6\% better with NA-RP than with SLB.
It works best with small $N_{\text{victim}}$, smallest $N_{\text{steal}}$, large $T_{\text{interval}}$, and largest $P_{\text{local}}$.
NA-RP brings more tasks to the same core and NUMA zone.
Most steal requests are successful.

%Stras and Sort
Both \textbf{STRAS} and \textbf{Sort} perform around 4$\times$ better with NA-RP than that with SLB.
They perform best with the largest $N_{\text{victim}}$ and $N_{\text{steal}}$, fully NUMA-local and small $T_{\text{interval}}$.
With SLB, most tasks are executed in a remote NUMA zone.
NA-RP successfully co-locates around 711K tasks for \textbf{STRAS}, and around 2.7M tasks for \textbf{Sort} to NUMA-local nodes: 85\% and 90\% of the total, respectively.
\textbf{STRAS} allocates large arrays in each task, bringing more tasks to the same core and NUMA zone can utilize faster memory access.
% Sort and Stras
Both \textbf{Sort} and \textbf{STRAS} allocate the memory space in an interleaved NUMA policy for the array before the OpenMP parallel region.
Their large task sizes result in early runtime under-utilization for most workers, so NA-RP takes place before SLB can do much work.
Therefore, many workers have requests to redirect tasks to NUMA-local nodes, leveraging multi-level cache and data proximity to reduce the latency and significantly increase performance.

% Align
\textbf{Align} does not have any steals with NA-RP and has negligible difference in performance with different parameters. 
\textbf{Align} uses the \emph{single} OpenMP construct where only one thread is responsible for creating all tasks in a loop.
These tasks are distributed by the default SLB and only the thread that runs the \emph{single} construct is able to redirect tasks with NA-RP.
\textbf{Align}'s task sizes follow a normal distribution, with the majority falling within the range of 10\textsuperscript{6}--10\textsuperscript{7} cycles.
Most of \textbf{Align}'s cycles are spent aligning the same set of limited-size protein sequences, which fit comfortably in the cache and therefore do not generate a significant communication between cores and main memory.

\subsubsection{NUMA-aware Work Stealing (NA-WS)}
\label{subsubsec:eval-ws}

All applications exhibit performance improvement with NA-WS, albeit with different configurations, as shown in \autoref{tab:combined-settings}.

\textbf{Fib} achieves negligible (max 1.3\%) performance improvement.
NA-WS performs best with the smallest $N_{\text{victim}}$ and $N_{\text{steal}}$, moderate $T_{\text{interval}}$, and uses only NUMA-local nodes.
\textbf{Fib}'s serial task graph means that there are few opportunities for parallelism and thus task stealing.
More self-executed tasks means more opportunities to utilize first-level cache.
NA-WS can take advantage of such opportunities because it can bring pushed tasks back to where they are created, thus increasing self-executed tasks.
However, over 95\% of tasks are already immediately executed on the same core when using SLB and therefore there is little room for improvement.

\textbf{NQueens} uses similar parameter settings as \textbf{Fib}, with larger $N_{\text{victim}}$ leading to the best performance improvement (9.8\%). 
The statistics indicate that NA-WS successfully steals 1.3M tasks. As a result, 3.7M tasks are subsequently created and executed on the same core compared to SLB.

\textbf{FFT} performs best (16.5\%) when $T_{\text{interval}}$ and $P_{\text{local}}$ are the same as \textbf{NQueens}, but with the largest $N_{\text{victim}}$ and largest $N_{\text{steal}}$.
From the statistics, NA-WS brings many more tasks back to its creators, with 1.7M self-executed tasks compared to 203K with SLB.
It also effectively moves 19.5M tasks to mitigate load imbalance.

\textbf{FP} achieves 2.8$\times$ performance improvement with moderate $N_{\text{victim}}$, the largest $N_{\text{steal}}$ and $T_{\text{interval}}$, and fully NUMA-local.
\textbf{FP} is a heavily imbalanced program with SLB and contains many tasks between 10\textsuperscript{3}--10\textsuperscript{6} cycles.
The performance gain is primarily due to NA-WS mitigating load imbalances while keeping many tasks local.

\textbf{Health} achieves the best performance improvement (32.5\%) with small $N_{\text{victim}}$, the largest $N_{\text{steal}}$, smallest $T_{\text{interval}}$, and a $P_{\text{local}}$ of 0.5.
From the statistics, NA-WS moves many tasks to its creators, with 33.5M self-executed tasks compared to only 658.2K with SLB.

\textbf{UTS} performs best (36.4\%) with small $N_{\text{victim}}$, largest $N_{\text{steal}}$ and $T_{\text{interval}}$, and fully NUMA-local.
Though NA-WS does not improve task locality, it actively moves 48.9M tasks to mitigate load imbalance.

\textbf{STRAS} achieves the best performance improvement (95\%) with small $N_{\text{victim}}$, largest $N_{\text{steal}}$, and smallest $T_{\text{interval}}$ and $P_{\text{local}}$.
Its performance gain is primarily due to moving more tasks to the same core and NUMA zone. 

\textbf{Sort} improves the most (76.9\%) with largest $N_{\text{victim}}$ and $N_{\text{steal}}$ and smallest $T_{\text{interval}}$ and $P_{\text{local}}$. 
It significantly increases the number of NUMA-local and self-executed tasks to exploit data locality.

\textbf{Align} performs best (9.6\%) with moderate $N_{\text{victim}}$, small $N_{\text{steals}}$, large $T_{\text{interval}}$ and fully NUMA-local.
Although it works well with SLB due to the application characteristics mentioned in \autoref{subsubsec:eval-ws}, NA-WS mitigates imbalance by moving 861K tasks.

In general, the lock-less NA-WS strategy can effectively mitigate load imbalance. With the proper settings, it can promote task locality by increasing the number of local and self-executed tasks.
It tends to steal tasks back to where they are created, which can benefit performance for both extremely fine-grained and coarser-grained tasks.
Although the performance for applications with larger task sizes is less than the NA-RP strategy, NA-WS achieves at least minimal performance improvement across all types of applications we tested.
Therefore, we conclude that NA-WS is a well-rounded and less sensitive work stealing approach that is most likely to achieve better performance.

\section{Application in Blockchain Consensus: Proof-of-Space (PoSp)}
\label{sec:application}

We investigate how our methods can improve the performance of a blockchain application using the Proof-of-Space (PoSp) consensus algorithm~\cite{Dziembowski2015,Cohen2019}. PoSp is an alternative to Proof-of-Work (PoW) that is less computationally expensive and reduces energy consumption. 
PoSp transforms a compute-intensive PoW problem into a data-intensive/storage problem where cryptographic puzzles are recorded in a persistent storage medium, later organized in order to be efficiently retrieved. PoSp uses the BLAKE3 cryptographic hashing algorithm~\cite{Oconnor2019BLAKE3} (over SHA-256) due to its excellent performance on a wide range of hardware. We extend a PoSp implementation that is written in C and using OpenMP~\cite{OpenMP5.0} as the primary mechanism to extract parallelism from the generation and storage of the cryptographic puzzles.

The PoSp implementation uses task-based parallelism in OpenMP to dynamically fill buckets with cryptographic puzzles until all buckets are full.
Task-based parallelism provides a flexible way to express irregular or dynamic parallel workloads that traditional loop-based parallelism may not handle efficiently. 
This is particularly useful when the workload is unbalanced or when the order in which tasks become ready to run is not strictly sequential, which is typically the case for PoSp. PoSp can be configured with number of threads, memory size, and batch size. The batch size determines the number of cryptographic puzzles to be generated in a single task. Small batch sizes can be inefficient if the runtime does not efficiently support small tasks, while large batches can be sub-optimal due to load imbalance.

% Vaultx Performance improvement
\begin{figure}[!ht]
    \centering
    \includegraphics[width=\columnwidth,trim=2mm 3mm 2mm 2.5mm,clip]{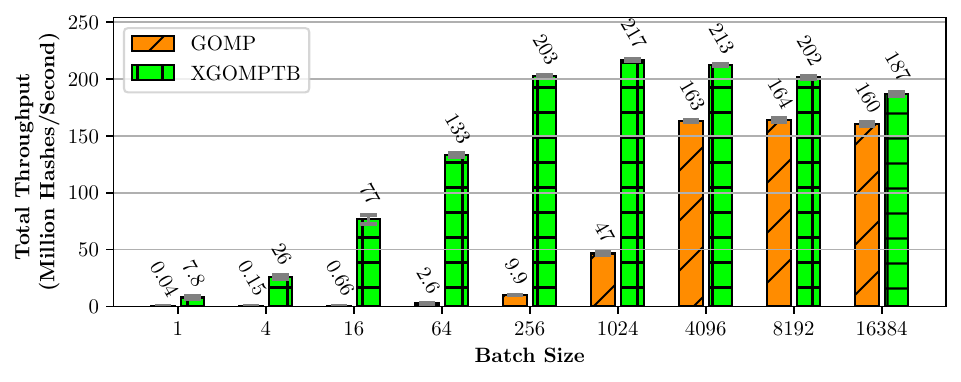}
    \caption{PoSp throughput comparing GOMP and XGOMPTB as the batch size is increased on 192-cores (higher is better).}
    \label{fig:vaultx-improve}
\end{figure}

The state-of-the-art production PoSp blockchains (e.g., Chia~\cite{ChiaGreenPaper2024}) use a minimum of $2^K$ (where $K=32$) cryptographic puzzles that are stored in a single file~\cite{Cohen2019, ChiaKSizes}. 
The PoSp we evaluated is made up of $2^K$ cryptographic puzzles, where each puzzle is composed of a 28\,Bytes BLAKE3 hash and a 4\,Bytes nonce value.
We conducted a parameter sweep (see \autoref{fig:vaultx-improve}) comparing XGOMPTB and GOMP on our 192-core system with increasing batch size. 
Using a batch size of 1, XGOMPTB achieves 7.8 megahashes per second (MH/s), compared to 0.04 MH/s for native GOMP, a 195$\times$ improvement. A batch size of 1 yields relatively small tasks, stressing the runtime’s ability to process a high throughput of tasks per second. XGOMPTB achieves 7.8\,M tasks per second while GOMP only achieves 40\,K tasks per second. The significant throughput reduction for GOMP comes from shared locks in the OpenMP runtime. The best performance is achieved for a batch size of 1024 for XGOMPTB with 217\,MH/s, compared to 164\,MH/s for GOMP with an 8192 batch size, a 32\% speedup. Batch sizes that are too large can yield lower performance due to load imbalance that leads to underutilized resources.

\section{Performance Tuning}
\label{sec:performance-tunning}

Our experiments reported in \autoref{sec:eval-dlb} show that different applications react differently to the choice of DLB and are sensitive to parameter choices.
For example, all applications are relatively sensitive to $N_{\text{victim}}$, $N_{\text{steal}}$, and $P_{\text{local}}$.
 
To further study the relationships between the performance and parameters, we categorize applications into five sizes based on per-task \emph{rdtscp} cycles $S_{\text{task}}$.
We then define steal size $S_{\text{steal}}$ in \autoref{eq:surface}, where $N_{\text{victim}}$ is the number of victims per request, $N_{\text{steal}}$ is the number of steals per victim, and $T_{\text{interval}}$ is a timeout counter for the worker before it retries the next round of requests. 
Since applications lack sensitivity to $T_{\text{interval}}$, we reduce its influence by applying a $\log$ function.

\begin{equation}
    S_{\text{steal}}= \frac{N_{\text{steal}} \times N_{\text{victim}}}{\log_{10}{T_{\text{interval}}}}\label{eq:surface}
\end{equation}

We show in Figs.~\ref{fig:ws-surface-A} and~\ref{fig:ws-surface-B} the performance improvement of both DLBs with 3D triangular surface plots.
In each figure, the Z-axis is the performance improvement over XGOMPTB.

\autoref{fig:ws-surface-A} shows that when using NA-RP, applications with task size $<$10\textsuperscript{2} cycles suffer performance degradation.
For 10\textsuperscript{2}--10\textsuperscript{4} cycles, we see little performance improvement.
For $>$10\textsuperscript{4} cycles, performance generally benefits from larger steal sizes.
For the largest tasks, NA-RP works exceptionally well, with $4\times$ performance improvement.

\begin{figure}[!ht]
    \centering
    \includegraphics[width=0.9\columnwidth,trim=2mm 0 4mm 4mm,clip]{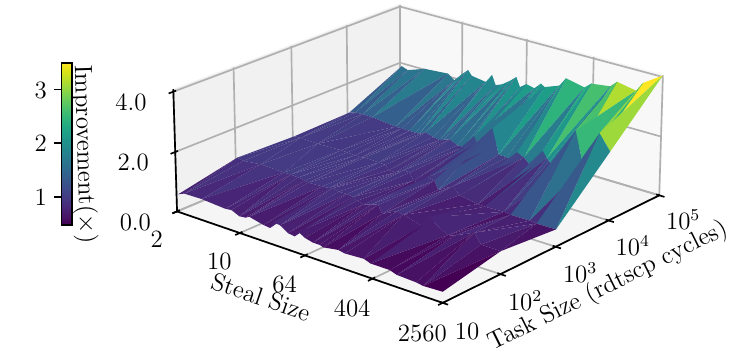}
    \caption{NA-RP (redirect-push) approach performance improvement ($\times$) over XGOMPTB as a function of task size and steal size (both log scale).}
    \label{fig:ws-surface-A}
\end{figure}

\autoref{fig:ws-surface-B} shows that NA-WS is less sensitive to configuration and leads to performance degradation only for applications with small tasks ($<$10\textsuperscript{3}) and large steal size.
As task size increases, application performance generally increases, with the best improvement achieved for the largest task and steal sizes. 
Generally, applications with larger tasks benefit more from larger steal sizes; applications with small tasks should use small steal sizes.

\begin{figure}[!ht]
    \centering
    \includegraphics[width=0.9\columnwidth,trim=2mm 0 4mm 4mm,clip]{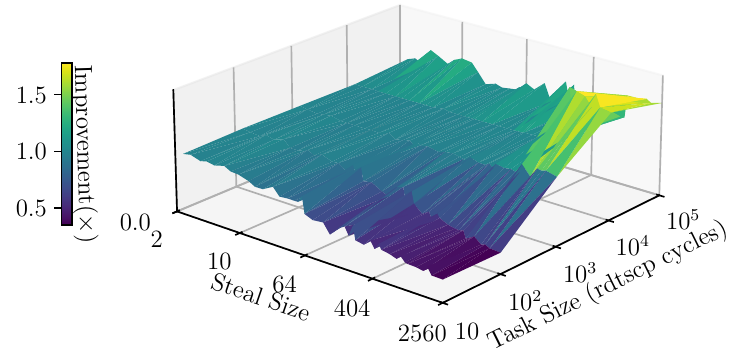}
    \caption{A-WS (work-stealing) approach performance improvement ($\times$) over XGOMPTB as a function of task size and steal size (both log scale).}
    \label{fig:ws-surface-B}
\end{figure}

% \begin{table*}[t]
% \caption{Optimal DLB settings for different task sizes}
% \centering
% \begin{tabular}{|l|c|c|c|c|c|}
% \hline
%  Size category      & Small     &Small-Medium      &Medium     &Medium-Large   &Large       \\
% \hline
% Task size in rdtscp cycles    &10\textsuperscript{1}--10\textsuperscript{2}       &10\textsuperscript{2}     &10\textsuperscript{3}     &10\textsuperscript{3}--10\textsuperscript{4}     &$>$10\textsuperscript{4}   \\ 
% \hline\hline
% Best DLB choice     & Work Stealing     &Work Stealing     &Work Stealing     &Work Stealing    & Redirect Push \\
% \hline
% Recommended (NUMA-local probabilty ($P_{\text{local}}$)             &100\%          &100\%      &100\%      &3--50\%         &3--12\% \\
% \hline
% Recommended steal size ($S_{\text{size}}$) &10\textsuperscript{0}--10\textsuperscript{1}  &10\textsuperscript{1}--10\textsuperscript{2}    &10\textsuperscript{2}--10\textsuperscript{2.5}    &10\textsuperscript{2.5}--10\textsuperscript{3}   & $>$10\textsuperscript{3}  \\
% % Recommended $S_{\text{size}}$ (Steal Size) &Small  &Small-Medium    &Medium    &Medium-Large   &Large  \\
% % Steal Size                  &Minimum        &Small      &Max        &Large-Max      &Max  \\ 
% \hline
% \end{tabular}
% \label{tab:guidelines}
% \end{table*}

\begin{table}[t]
\caption{Optimal DLB settings for different task sizes.}
\centering
\resizebox{\columnwidth}{!}{%
\begin{tabular}{|l|c|c|c|c|c|}
% \hline
 % Size     & Small     &Small-Medium      &Medium     &Medium-Large   &Large       \\
\hline
% \bottomrule
$S_{\text{task}}$ (rdtscp)    &10\textsuperscript{1}--10\textsuperscript{2}       &10\textsuperscript{2}     &10\textsuperscript{3}     &10\textsuperscript{3}--10\textsuperscript{4}     &$>$10\textsuperscript{4}   \\ 
\hline
% \hline
Best DLB & WS     &WS     &WS     &WS    & RP \\
\hline
Best $P_{\text{local}}$             &100\%          &100\%      &100\%      &3--50\%         &3--12\% \\
\hline
Best $S_{\text{steal}}$ &10\textsuperscript{0}--10\textsuperscript{1}  &10\textsuperscript{1}--10\textsuperscript{2}    &10\textsuperscript{2}--10\textsuperscript{2.5}    &10\textsuperscript{2.5}--10\textsuperscript{3}   & $>$10\textsuperscript{3}  \\
% Recommended $S_{\text{size}}$ (Steal Size) &Small  &Small-Medium    &Medium    &Medium-Large   &Large  \\
% Steal Size                  &Minimum        &Small      &Max        &Large-Max      &Max  \\ 
\hline
\end{tabular}
}
\label{tab:guidelines}
\end{table}

Locality is also an important contributor to performance. 
For NA-RP, fully local redirection yields the best performance across all benchmarks (see \autoref{tab:combined-settings}) because it optimizes memory access latency by pushing tasks to adjacent workers. 
Because NA-RP pushes more tasks away, it is the best choice to send tasks to the local node, which can also utilize multi-level cache. 
For NA-WS, extremely fine-grained tasks perform best with fully NUMA-local steal, to optimize the overall memory access latency. 
For larger tasks, smaller $P_{\text{local}}$ performs better because work stealing can return pushed tasks to where they are created, allowing them to utilize high-level cache which optimizes overall latency.

In summary, different combinations of strategies and parameters suit different application characteristics. 
Applications with extremely fine-grained tasks benefit from smaller steal sizes on the NA-WS approach; full NUMA-local probability should be used. 
Applications with larger tasks benefit from larger steal sizes. 
NA-RP with maximum steal size and fully NUMA-local is particularly suitable for tasks of $>$10\textsuperscript{4} cycles. 
We summarize in \autoref{tab:guidelines} guidelines for selecting parameters for the best performance and
show in \autoref{fig:best-all-large} results obtained for the five applications when using these guidelines. We use as input arguments,
\textbf{Fib}: 42, \textbf{NQueens}: 16, \textbf{FFT}: 536M, \textbf{FP}: 20, \textbf{Health}: xlarge, \textbf{UTS}: small, \textbf{STRAS}: 4096 (Y=16), \textbf{Sort}: 1B, and \textbf{Align}: 2000.

\begin{figure}[!ht]
    \centering
    \includegraphics[width=\columnwidth,trim=2.5mm 3mm 2mm 2.5mm,clip]{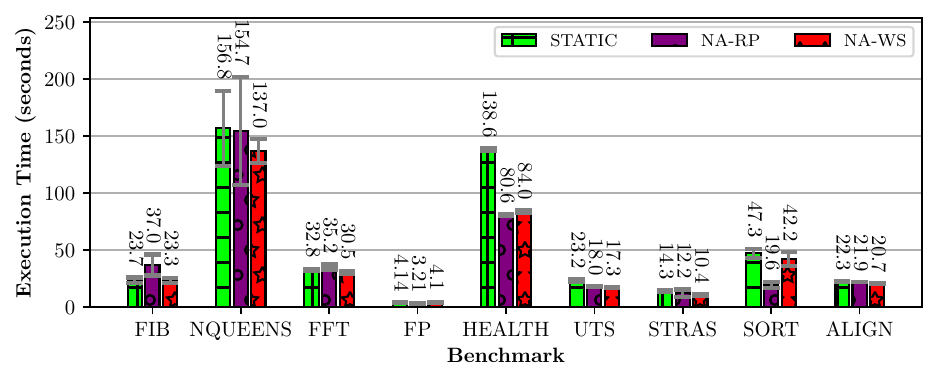}
    \caption{XGOMP, NA-RP, and NA-WS performance comparison on BOTS when following the guidelines of \autoref{tab:guidelines}.}
    \label{fig:best-all-large}
\end{figure}

% RELATED WORK

\section{Related Work}
\label{sec:related-work}

%\noindent
\textbf{Parallel runtime systems} 
%OpenMP, Cilk, libfork, tbb, mpi, charm++
such as OpenMP \cite{miscname4}, Charm++ \cite{kale1993charm}, and Swift/T \cite{wozniak2013swift} use concurrent queues to share data between threads or processes.
Charm++ goes further and supports lock-free queues to demonstrate performance improvements \cite{10.1145/1838574.1838586}.
Regarding locks, researchers have investigated contention management in thread-safe MPI libraries \cite{10.1145/3275443} and the use of abort locking \cite{10.1145/3155284.3018768} to improve performance.
We aim to achieve better performance with finer granularity and task decomposition than these solutions.

Some classical with-locks task-based runtimes (e.g., OmpSs~\cite{ompss12}, PaRSEC~\cite{parsec}, StarPU~\cite{starpu}) have made efforts to improve data locality. 
For example, OmpSs and XKaapi~\cite{xkaapi15} rely on work-stealing for load balancing.
XKaapi also provides a lower bound on the number of data accesses required by the scheduler~\cite{workstealing_locality02}.
Legion~\cite{legion_overview_dup} allows users to specify locality explicitly using data regions, and provides a data mapping strategy to ensure that data are only moved when needed.
Some of StarPU's scheduling strategies focus on data reuse and task stealing to increase the performance of linear algebra applications~\cite{gonthier-ipdps, gonthier:hal-03290998}.
These solutions are orthogonal to our research: We focus on removing the synchronization cost of barriers and thus cannot use similar techniques that require regular synchronization and updating of current system state.

%\noindent
\textbf{Dynamic Load balancing:}
% Theoretical paper on random NA-WS, hierarchical work stealing, scabale work stealing, adws
Several papers have proposed load
balancing mechanisms~\cite{10.1145/1654059.1654113, guo2009work}.
Quintin et al. proposed hierarchical work stealing to exploit data
locality to achieve speedup over classical work stealing algorithms~\cite{10.1007/978-3-642-15277-1_21}. 
Shiina et al. addressed the issue of data locality by making scheduling deterministic~\cite{10.1145/3295500.3356161}.
In order to balance load efficiently, these all relied on synchronization mechanisms.
In our work, we explore lock-less techniques to achieve comparable dynamic load balancing mechanisms.

\textbf{Lock-less runtime systems:}
XQueue~\cite{nookala_enabling_2021} demonstrated the benefits of a lock-less, task-oriented runtime in OpenMP.  
Recent work on XQueue~\cite{NOOKALA2024444} introduced work-stealing, but focused on simply redirecting a newly spawned task to another thread. Results indicated a need to develop more sophisticated strategies and also consider NUMA-awareness, as we explore here.
Among other things, in this paper, we integrate XQueue into GNU-OpenMP, introduce a tree barrier, and propose sophisticated NUMA-aware work-stealing strategies, which dynamically migrate tasks from a victim to a thief thread.

% DISCUSSION AND FUTURE WORK

\section{Discussion and Future Work}
\label{sec:discussion}

GNU OpenMP, although part of the mainstream compiler infrastructure GCC, is unable to harness modern high-performance CPUs with many cores, due to its %conservative design of using excessive 
use of locks and atomic operations. 
Our work improves GNU OpenMP %into a high-performance parallel computing library 
performance by integrating a novel lock-less concurrent data structure, XQueue, and an efficient distributed tree barrier to achieve up to 1522.8$\times$ improvement compared to the original GNU OpenMP.
We further improve performance through the use of dynamic load balancing strategies, demonstrating that with optimal settings, lock-less dynamic load balancing can achieve 4$\times$ more performance improvement compared to using static load balancing. 

Our experiments show that good parameter choices are dependent on application characteristics. 
We provide guidance to help practitioners select parameter values based on their applications.
In future work, we will decompose application characteristics to automate the selection of good settings.

\section*{Acknowledgments}
This work is supported in part by the NSF CNS-1730689 CRI award (Mystic Testbed~\cite{orhean2019mystic}), as well as the NSF OAC-2107283 and OAC-2107548 awards. 
%The team is also thankful to the anonymous reviewers for providing valuable feedback for improving this work.

\balance
\bibliographystyle{IEEEtran}
\bibliography{main}

\end{document}